\documentclass[epj]{svjour}
\usepackage{graphicx}
\pdfoutput=1
\begin{document}
\title{Generation of slow intense optical solitons in a resonance photonic crystal }

\author{
Igor V. Mel'nikov\inst{1}
\thanks{\emph{E-mail:} igor.melnikov@utoronto.ca}%
, Anton Knigavko\inst{1,2},
J. Stewart Aitchison\inst{3}, and Clark A. Merchant\inst{3}
}                     
%
%
\institute{
High Q Laboratories, Inc., 2 Gledhill Cres., Hamilton, Ontario L9C 6H4, Canada
\and
Department of Physics, Brock University,
500 Glenridge Ave., St.~Catharines, Ontario L2S 3A1, Canada
\and
Department of Electrical and Computer Engineering, University of Toronto,
10 King's College Rd., Toronto, Ontario M5S 3G4, Canada
}
\date{Received: date / Revised version: date}
%
\abstract{ We demonstrate interesting and previously unforeseen
properties of a pair of gap solitons in a resonant photonic crystal
which are predicted and explained in a physically transparent
form using both analytical and numerical methods. The most important
result is the fact that an oscillating gap soliton created by the
presence of a localized population inversion inside the crystal can
be manipulated by means of a proper choice of bit rate, phase
and amplitude modulation. Developing this idea, we are able to
obtain qualitatively different regimes of a resonant photonic
crystal operation. In particular, a noteworthy observation is that
both the delay time and amplitude difference must exceed a certain level
to ensure effective control over the soliton dynamics.
} 
\titlerunning{Generation of slow intense optical solitons ...}
\authorrunning{I.~V. Mel'nikov, A. Knigavko, J.~S. Aitchison, and C.~A. Merchant
}
\maketitle
\section{Introduction}
\label{intro}
There has been an extensive amount of research focused on the study on nonlinear
periodic systems and in particular those described by the sine-Gordon equation.
This is primarily due to a large and diverse range of physical phenomena which
can be described by the soliton solutions of this equation. Resonance photonic crystals (RPhC)
represent one such system where the sine-Gordon equation can be used to describe
the propagation of light. Periodicity in photonics results in bandgaps,
or regions in time or space, where the propagation of energy is forbidden \cite{ref1,ref2,ref3}.
The addition of nonlinearity to the system allows the possibility of locally detuning
these band gaps with the result that energy can propagate in the form of a solitary wave,
or gap soliton (GS) \cite{ref4,ref5,ref6,ref7}. Such gap solitons have been experimentally
observed in fiber Bragg gratings \cite{ref8} and more recently in AlGaAs waveguides
\cite{ref9,ref10,ref11}.
These experiments relied on the non-resonant, Kerr nonlinearity to locally detune
the stop-bands associated with the periodic structure.

A recent resurgence of interest in photonics crystals with resonant nonlinearities is
due to the advances in the manufacturing of periodic Bragg semiconductor nanostructures
such as InGaAs/GaAs multiple quantum wells and rare-earth doped AlGaAs/GaAs structures
\cite{ref12,ref13,ref14},
and understanding their linear and nonlinear optical properties \cite{ref15,ref16,ref17}.
This suggests that the building blocks of future all-optical processing systems,
such as those based on the propagation of GSs might be experimentally feasible
at moderate optical power levels of about 10 MW/cm$^2$. The inclusion of a gain or a local
population inversion in the system opens up a range of additional, novel nonlinear
effects \cite{ref18}.

Recently, we proposed an effective method for the control of speed of gap solitons
based on the defect associated with a weak linear excitation, or incoherent pump
inside an RPhC \cite{ref19,ref20,ref21,ref22}.
It turns out that a break in periodicity of the RPhC is
not necessary for slowing a gap soliton down and its trapping because the localized
field creates a potential that acts as a phase-sensitive trap. The GS interaction
at such defect was first studied to determine the gain implication for slow light
localization in the RPhC \cite{ref19}. More recent work was focused on the effects of
localized gain on the GS collision and comprehensive theory was elaborated \cite{ref20,ref21}.
Light trapping at the gain inside an RPhC with subsequent release due to the
collision with another GS have also been predicted and corresponding memory device
was also suggested \cite{ref22}.  In this paper, we look at the two separate effects
associated with a localized population inversion inside RPhCs:
higher-order soliton dynamics in the presence of such gain defects and generation
of the gain defect itself.
This paper is subdivided as follows. In Sect.~\ref{analytic}, the bi-directional
Maxwell-Bloch model is formulated and analytical results are given for the exact
GS interacting with the localized gain inside the RPhC.
Sect.~\ref{numeric} contains numerical results for the interaction of a GS train
with this inversion length. Sect.~\ref{discuss} discusses
possible experimental scenarios and presents conclusions.

\section{Analytical Formalism}
\label{analytic}

In order to explore the basic features of interaction of resonant gap solitons in
the presence of a localized gain or defect inside the photonic crystal we consider
the simplest cases where the defect is created in the middle of the RPC using
an incoherent pump with square hump-like spatial profile

\subsection{Coupled-Mode Equations}
\label{CMeq}

The 1D resonant photonic crystal is assumed to comprise of periodically
positioned thin layers of two-level atoms. The electric field that appears in the
Maxwell's equations and couples the two levels is taken in the form of a superposition
of the forward, $E^{+}$, and backward-propagating wave, $E^{-}$, as follows:
\begin{eqnarray}
\textstyle
\label{eq1}
  E(x,t) &=& \frac{1}{2}
    \big[ E^{+}(x,t)e^{i( k x - \omega t)} +  E^{-}(x,t)e^{-i(k x + \omega t)} \big].
\end{eqnarray}
The direction of propagation is normal to the layers. The incident radiation has
a wavelength $\lambda$ which exactly satisfies the Bragg condition $d = \lambda$,
where $d$ is the RPhC period. We also assume that the field described by Eq.~(\ref{eq1})
is tuned into the exact resonance with atomic absorption of the two-level system
(which can represent atoms, ions, or excitons). In this case the dynamics
of a two-level system, can be described by the generalized
Bloch equations as follows \cite{ref5,ref19,ref20,ref21,ref22,ref23}:
\begin{eqnarray}
\label{eq2a}
  P_t(x_i,t) &=& n(x_i,t) \big[
    \Omega^{+}(x_i,t)e^{i k x_i} +  \Omega^{-}(x_i,t)e^{-i k x_i}
  \big],\\
 \nonumber
  n(x_i,t) &=& -{\rm Re} \Big\{ P^*(x_i,t) \big[ \Omega^{+}(x_i,t)e^{i k x_i}
  \\
\label{eq2b}
  && \phantom{WWWWWWW}+  \Omega^{-}(x_i,t)e^{-i k x_i}
  \big] \Big\},
\end{eqnarray}
where $\Omega^{\pm} = 2\tau_c \mu E^{\pm}/\hbar$ with $E^{\pm}$ from Eq. (\ref{eq1}) and
$\tau_c = (8\pi \varepsilon T_I /3c\rho \lambda^2)^{1/2}$ is the cooperative time that
is the mean photon lifetime in the structure, $T_I$ is the excited state lifetime,
$\varepsilon$ is the dielectric constant of the medium,
$\rho$ is the density of the resonant atoms, $\mu$ is the transition matrix element,
$x_i$ is the position the i-th resonant layer, $c$ is the speed of light in vacuum,
and the subscript $t$ indicates the corresponding derivative.
In turn, the resonant polarization $P$ defines the dynamics of the optical field
$\Omega^{\pm}$ inside the medium via the set of the two-mode equations:
\begin{eqnarray}
\label{eq3}
  \Omega^{\pm}_t(x,t) \pm \Omega^{\pm}_x(x,t) &=&
  \sum_i e^{\mp i k x_i} P(x_i,t) \delta(x-x_i),
\end{eqnarray}
which after averaging over the structure period forms the closed set with
Eq.~(\ref{eq2a}) and Eq.~(\ref{eq2b}). This is shown here as:
\begin{eqnarray}
\label{eq4a}
\Omega^{\pm}_t \pm \Omega^{\pm}_x &=& P, \\
\label{eq4b}
P_t &=& n(\Omega^{+} + \Omega^{-}) \\
\label{eq4c}
n_t &=& - {\rm Re} \big\{ P^* (\Omega^{+} + \Omega^{-}) \big\},
\end{eqnarray}
where the subscripts $x$ and $t$ stand for the corresponding partial derivative along
the dimensionless time and propagation coordinate $t = t/\tau_c$ and $x = x/(c\tau_c)$.
As input values, we use the following boundary and initial conditions:
\begin{eqnarray}
\label{eq5a}
    \Omega^{+}_t(x=0,t) &=& \sum_{i=1}^{2} \Omega_i \;{\rm sech}[(t-t_i)/\tau_p], \\
\label{eq5b}
    \Omega^{-}(x=L,t) &=& 0, \\
\label{eq5c}
  \Omega^{\pm}_t(x,t=0) &=& 0, \\
\label{eq5d}
    P(x,t=0) &=& 0, \\
\label{eq5e}
  n(x,t=0) &=& \left\{\begin{array}{cc}
                        1,\quad & |x-L/2|\leq \ell/2, \\
                        -1,\quad & |x-L/2|\geq \ell/2,
                      \end{array}
  \right.
\end{eqnarray}
where $\tau_p$ is the normalized duration of the incident pulse.
Eqs.~(\ref{eq5a})--(\ref{eq5e}) together with Eqs.~(\ref{eq4a})--(\ref{eq4c})
fully define the system under considertion.

\subsection{Lagrangian  and mechanical analogy}
\label{analogy}

The invariable quantity of Eqs.~(\ref{eq4a})--(\ref{eq4c}) is accessible
via a transformation to the Bloch angle $\theta(x,t)$,
the sum $U = \Omega^{+} + \Omega^{-}$ and the difference $V = \Omega^{+} - \Omega^{-}$
fields, which leads to the following set of equations:
\begin{eqnarray}
\label{eq6a}
  U_t + V_x &=& -2 \sin \theta, \\
\label{eq6b}
  V_t + U_x &=& 0, \\
\label{eq6c}
  \theta_t &=& U.
\end{eqnarray}

An invariable quantity $\Gamma(x)$ associated with this set of equations can be obtained
by substituting Eq. (\ref{eq6c}) into Eq. (\ref{eq6b}) and and results in
\begin{eqnarray}
\label{eq7}
  \Gamma(x) &=& V(x,t) + \theta_x (x,t).
\end{eqnarray}

This conservation parameter is defined by the initial conditions and does not change
as the pulse propagates through the RPhC. However, it does describe the behavior of
the GS in the sense that setting it equal to zero provides a GS whereas, a non-zero value
of $\Gamma(x)$ leads to a lost of stability of the GS and can result in trapping of the GS inside
the RPhC. It is readily seen that the substitution of the invariant function described
by Eq. (\ref{eq7}) into Eq. (\ref{eq6a}) transforms it into the perturbed Sine-Gordon equation for the
Bloch angle $\theta$ shown as:
\begin{eqnarray}
\label{eq8}
  \theta_{xx} - \theta_{tt} &=& 2\sin \theta  + \Gamma_x .
\end{eqnarray}

In order to elucidate the interaction between the exact GS and defect mode that
is due to the localized gain, we apply the approach developed earlier in \cite{ref23}
where the coordinate of the GS center-of-mass $\zeta(t)$ can be derived from
a corresponding Lagrangian density:
\begin{eqnarray}
\label{eq9}
  L &=& \frac{1}{2} \theta^2_{t} - \frac{1}{2} \big(\theta_{x} - \Gamma \big)^2
  - (1-\cos \theta),
\end{eqnarray}
with corresponding density of the Hamiltonian
\begin{eqnarray}
\label{eq10}
  H &=& \frac{1}{2} \theta^2_{t} + \frac{1}{2}\theta^2_{x} - \Gamma \theta_{x}
  + \frac{1}{2} \Gamma^2 + (1-\cos \theta),
\end{eqnarray}
where the first four terms on the right-hand side define the energy density
$(U^2 + V^2)/2$ of the forward and backward waves inside the RPhC.

Assume now that the solution of Eq. (\ref{eq8}) can be approximated by that of the
non-perturbed sine-Gordon equation:
\begin{eqnarray}
\label{eq11}
  \theta(x,t) &=& 4 \arctan \left( \exp \left[ \frac{-x+\zeta(t)}{\sqrt{1-u^2(t)}}
  \right] \right),
\end{eqnarray}
where $\zeta(t)$ and $u(t)$ are the time-dependent coordinate of  the GS center and velocity,
correspondingly. Since the total energy of the localized solution is an invariable quantity,
the energy of the interaction of the soliton of Eq. (\ref{eq11}) with perturbation created by
the localized gain, is defined by the overlap integral \cite{ref20,ref21,ref22,ref23}:
\begin{eqnarray}
\label{eq12}
  \Phi &=& \frac{1}{4} \int_{-\infty}^{+\infty} d x \Gamma_x \theta_x,
\end{eqnarray}
which in turn describes the GS propagation across the RPhC with a length of the gain
inside as a classical motion of the particle with unit mass,
\begin{eqnarray}
\label{eq13}
  \zeta_{tt} &=& -\Phi_{\zeta}.
\end{eqnarray}

\begin{figure}
\includegraphics[viewport=50 220 500 550,width=.9\columnwidth,clip]%
{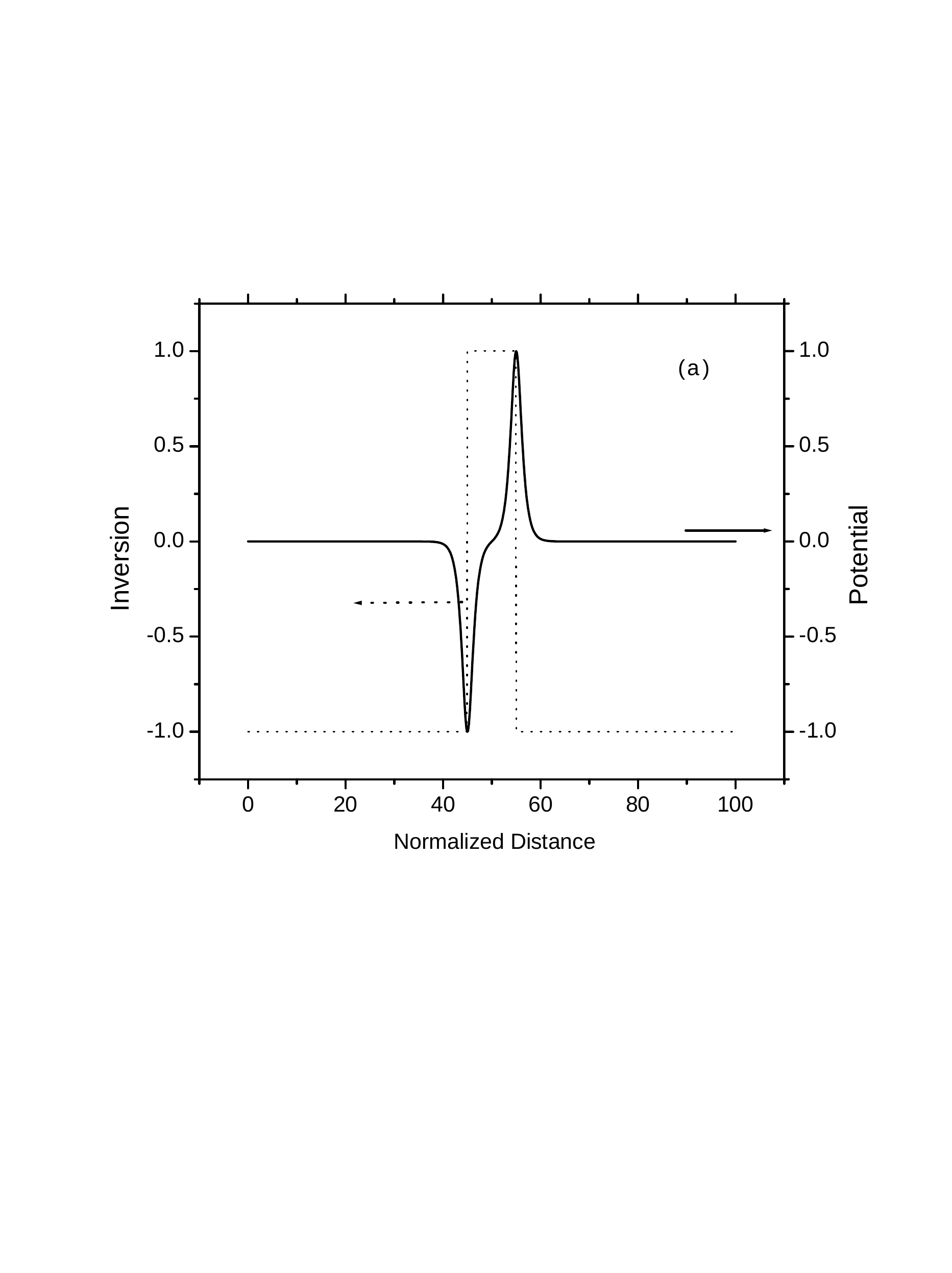}
\includegraphics[viewport=20 160 500 550,width=.9\columnwidth,clip]
{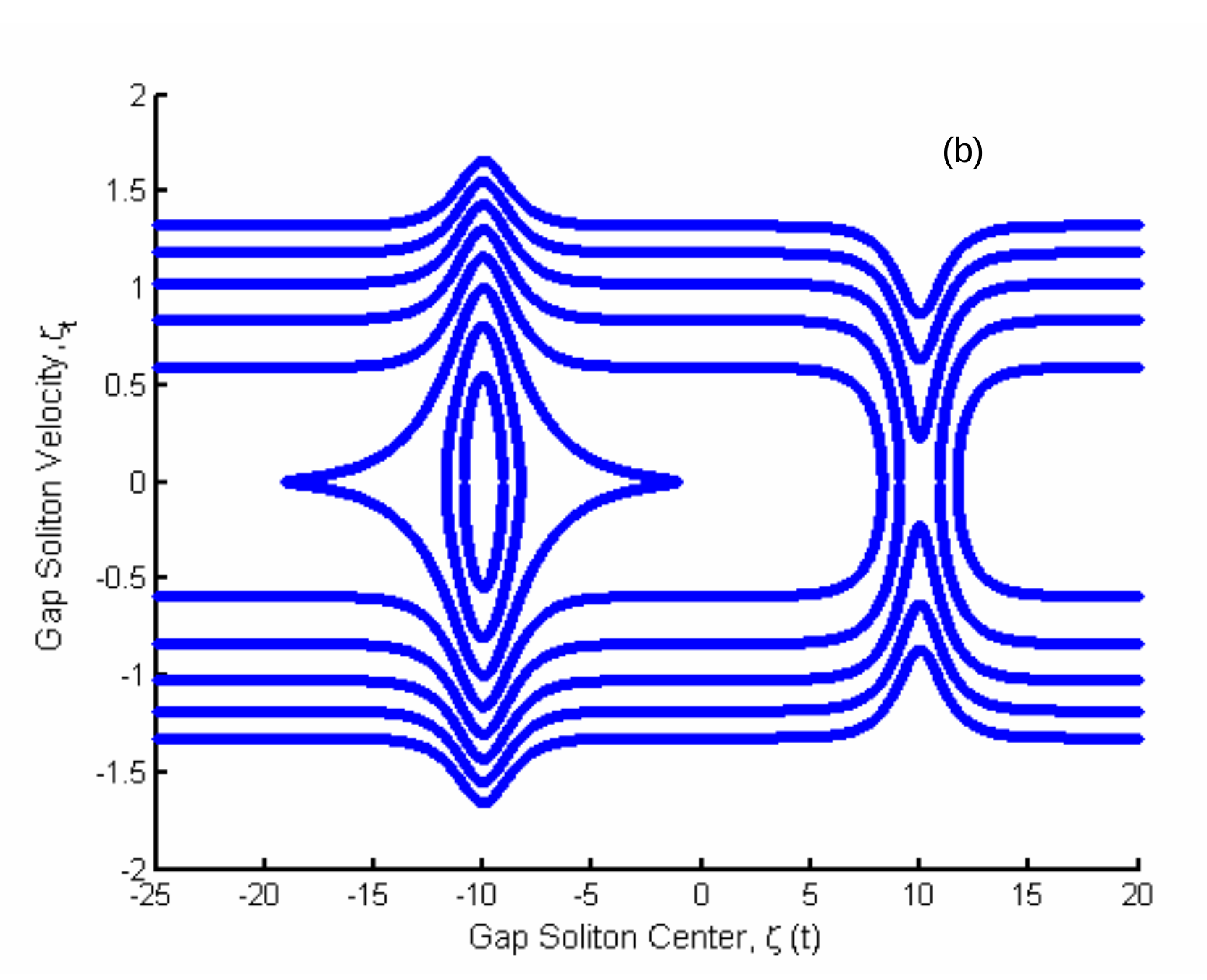}
\caption{
Profile of inversion $n$ (dotted line) and corresponding potential $\Phi$ (solid line)
created in the RPhC for the GS of Eq.~(\ref{eq11}) (a)
along with corresponding phase portrait (b);
notice existence of turning point, separatrix trajectory,
and closed aperiodic cycles.
}
\label{fig1}
\end{figure}

Our initial conditions of the gain localized in the middle of the RPhC, namely:
\begin{eqnarray}
\label{eq14}
  \theta(x,t=0) &=& \left\{\begin{array}{ccc}
                        0,\quad & 0 \leq x \leq (L-\ell)/2, \\
                        \pi,\quad & (L-\ell)/2 \leq x \leq (L+\ell)/2, \\
                        0,\quad & 0  \leq (L+\ell)/2 \leq x \leq L ,
                      \end{array}
  \right.
\end{eqnarray}
along with the absence of the propagating waves in the beginning,
generates the  following potential:
\begin{eqnarray}
\label{eq15}
 \Phi &\sim& - {\rm sech}\big[x-(L+\ell)/2\big] + {\rm sech}\big[x-(L-\ell)/2\big].
\end{eqnarray}
The profile of this potential and corresponding phase trajectories are depicted
in Fig. \ref{fig1}; it is readily seen that if the velocity of the particle (pulse) is
considerably above the potential zero level, it experiences some minor variations while
traversing the potential, or gain hump in our case. Lowering the velocity shifts
the phase trajectory into vicinity of the saddle point (not shown in Fig. 1b)
and the potential becomes repelling that implies a blockage of the GS free transport
across the RPhC. Next, if the velocity is small enough, the particle experiences
aperiodic oscillations inside the potential well on the left. That is, the slower
GS is trapped by the gain and oscillates around the gain center, it is unsteady
but stable. It is worth noticing that the energy gain of the GS that is due to
the gain of Eq. (\ref{eq14}) along with a modification of this gain by
the GS can change the dynamics of the interaction to a very large extend.

\begin{figure}
\includegraphics[viewport=100 280 500 600,width=.9\columnwidth,clip]
{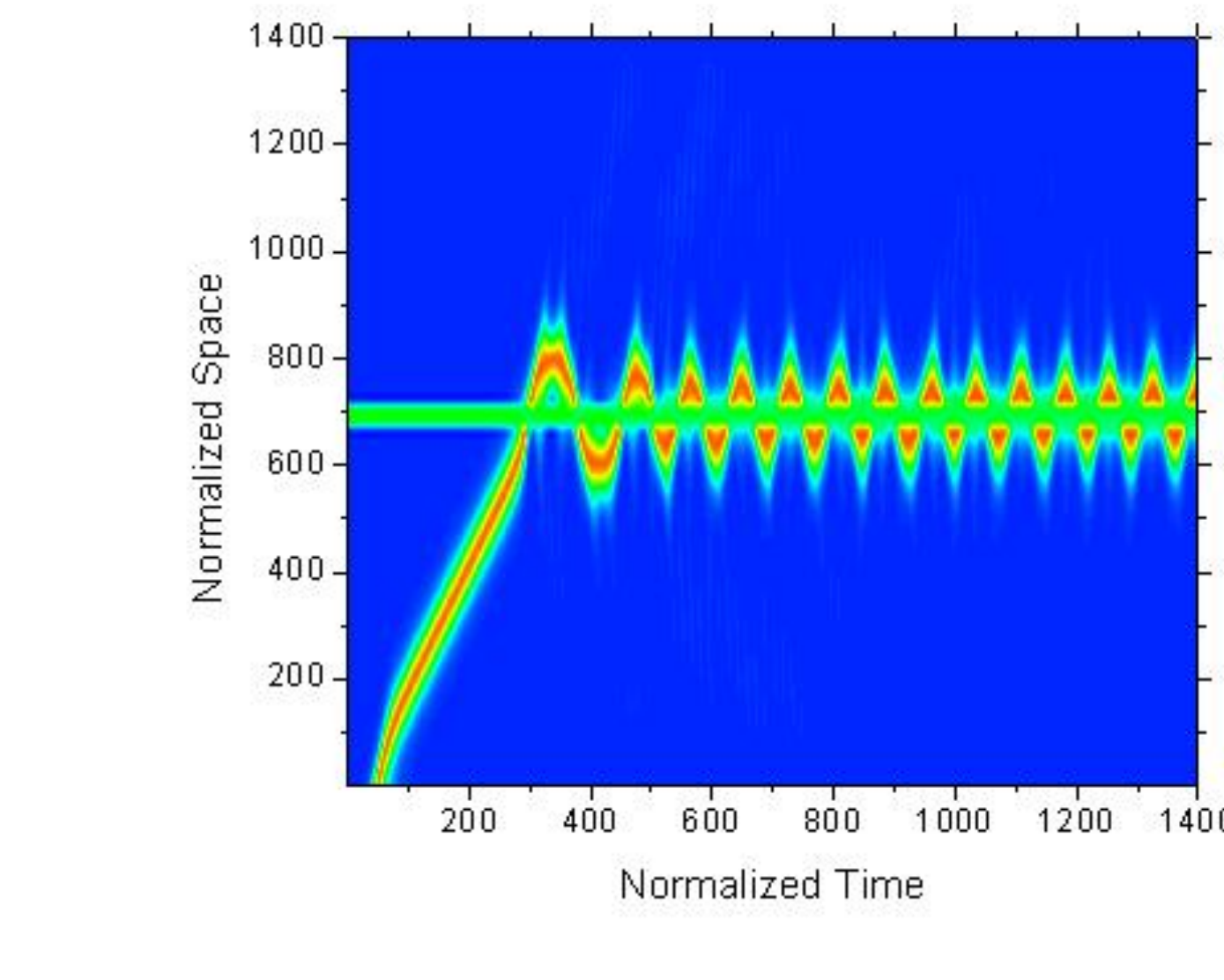}
\caption{ Localization and storage of a single $2\pi$ pulse inside
RPhC on an inversion induced defect; $\Omega_1= 1, t_1 = 50$,
$\Omega_2 = 0, \ell/L = 0.02$, where $\ell$ is the length of the
inversion and $L$ is the RPhC length; time and distance are
normalized to the coherence time and length, correspondingly. }
\label{fig2}
\end{figure}

\begin{figure}
\includegraphics[viewport=120 320 500 640,width=.9\columnwidth,clip]
{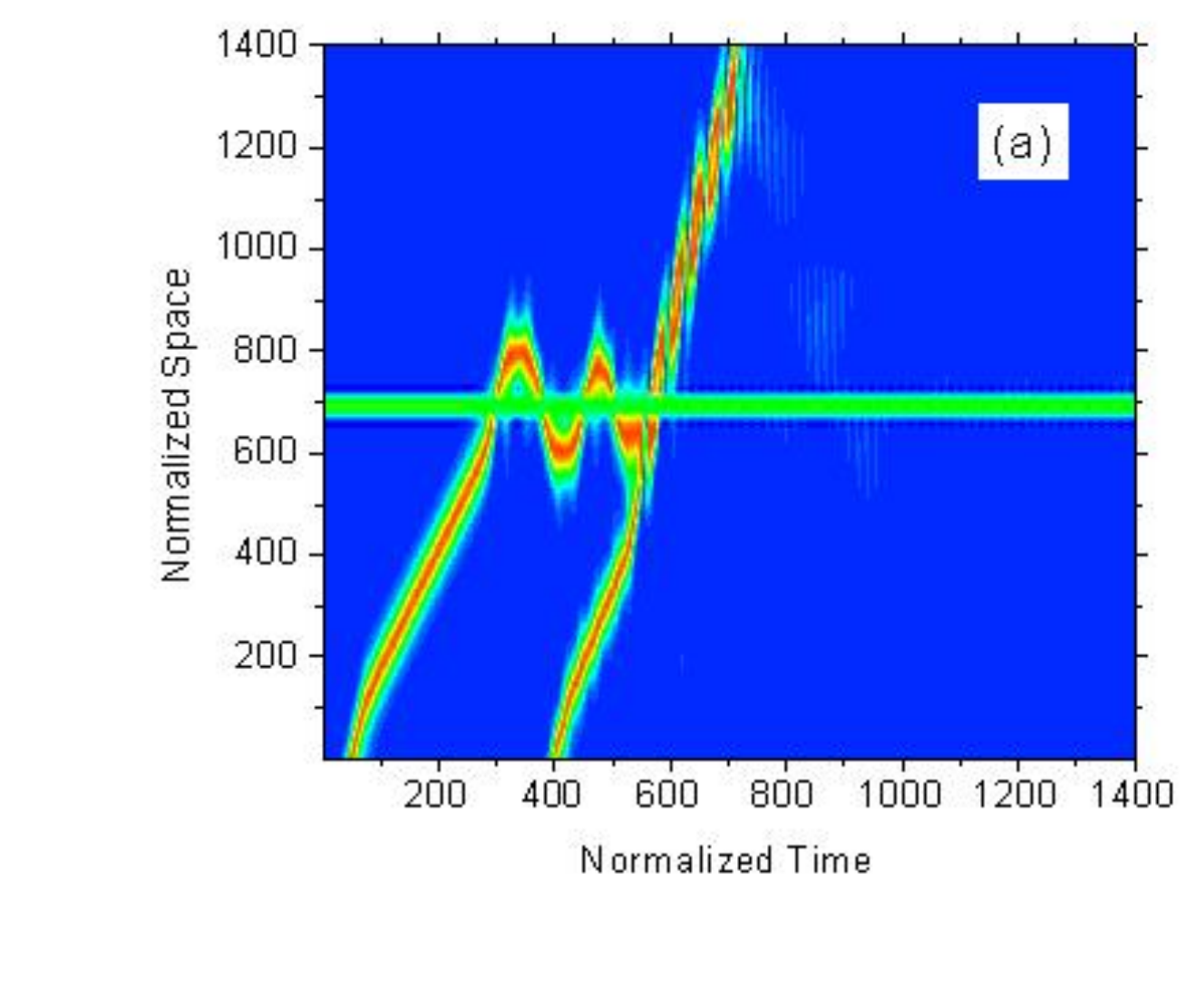}
\includegraphics[viewport=100 320 500 640,width=.9\columnwidth,clip]
{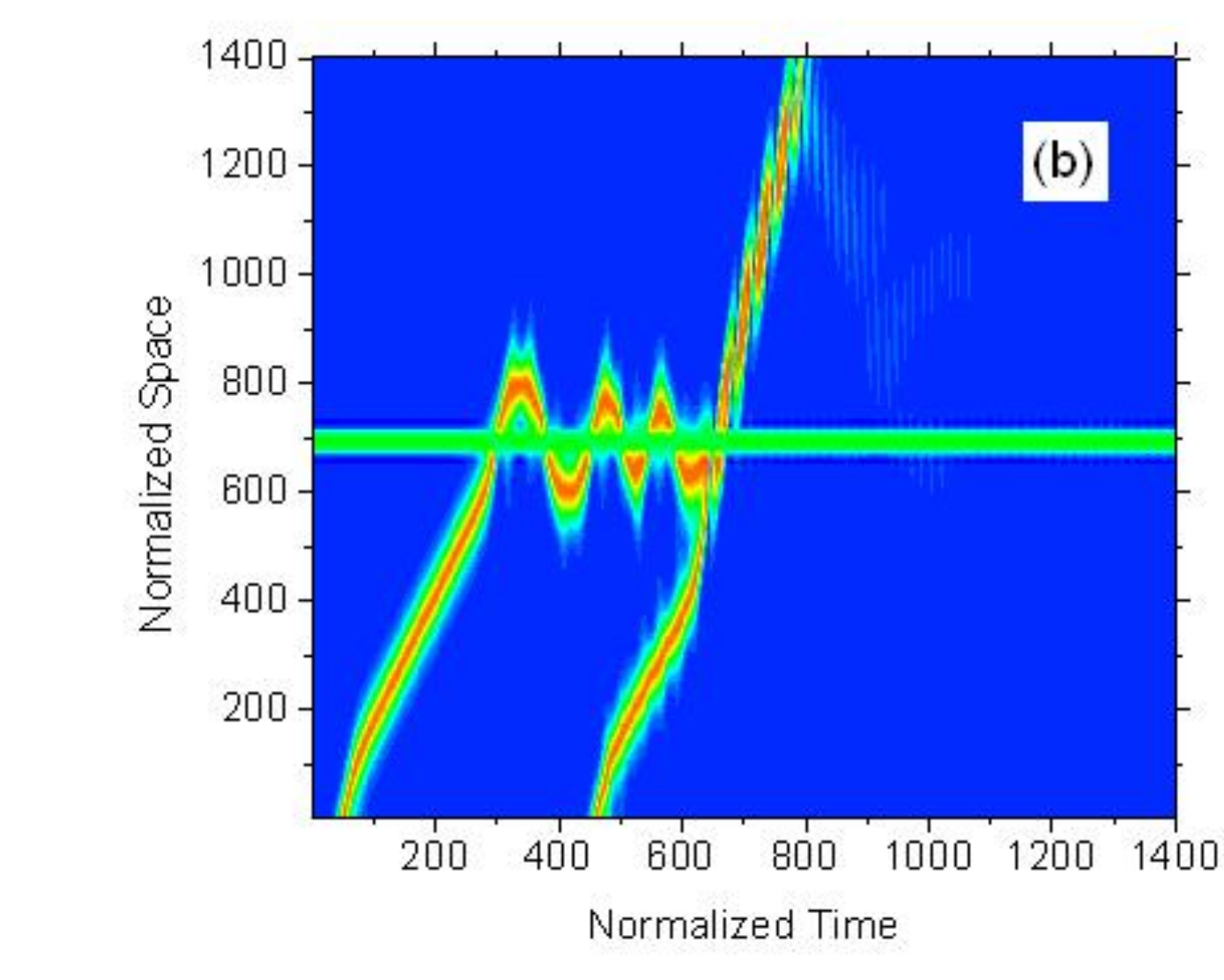}
\includegraphics[viewport=100 280 500 600,width=.9\columnwidth,clip]
{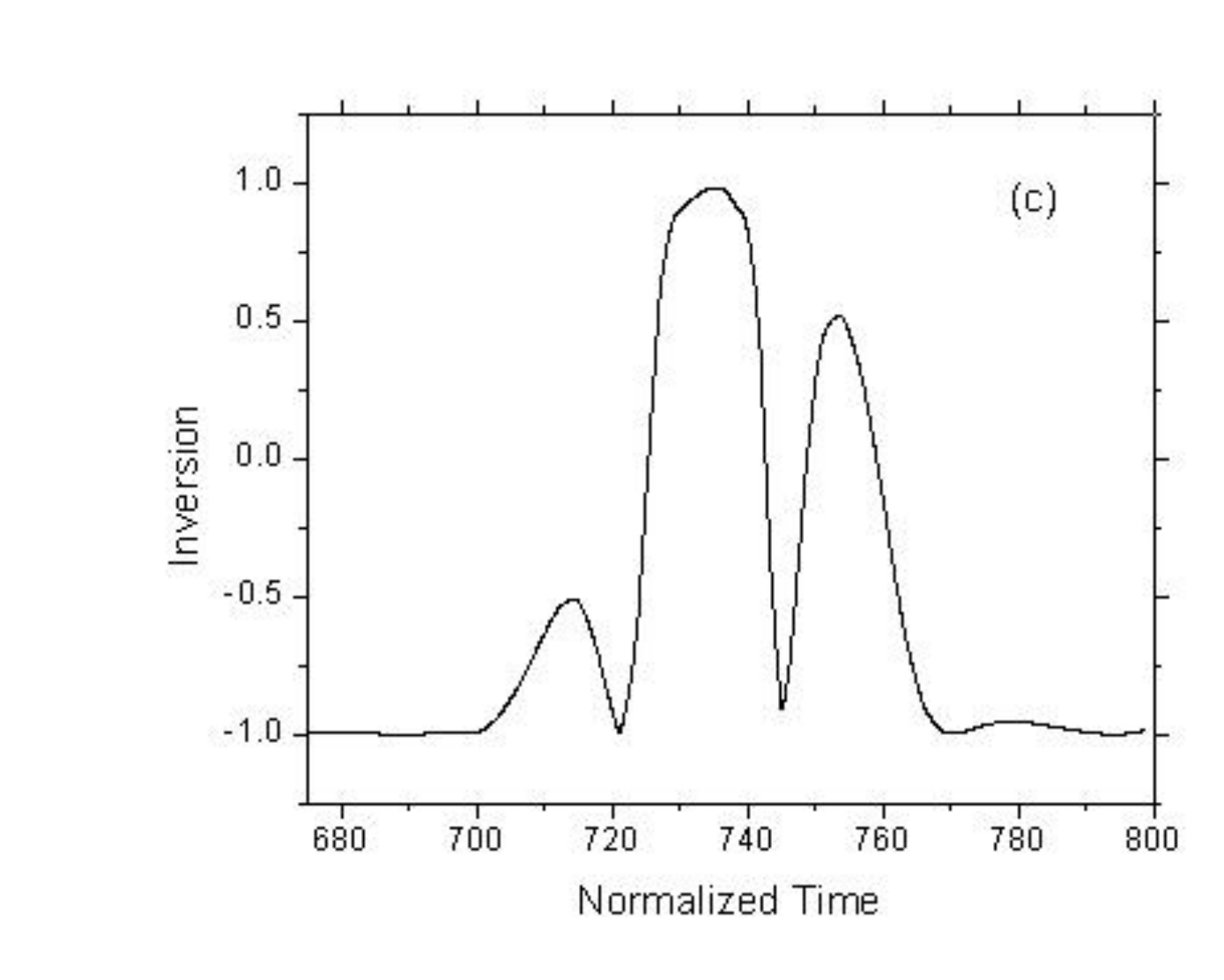}
\caption{ Release of a stored optical pulse by out-of-phase readout
$2\pi$ pulse and their merger after (a) 1.5 oscillations around the
defect (delay time $t_2 -t_I = 350$) and (b) 2.5 oscillations around
the defect defect (delay time $t_2 -t_I = 400$), $\Omega_1= 1 =
-\Omega_2$, $\ell/L = 0.02$; (c) breathing structure of the merger
pulse --- inversion scan at $x = 1100$ for (b). } \label{fig3}
\end{figure}

\section{Calculated Behavior and Discussion}
\label{numeric}

The analysis of the previous Section is now used to explore the behavior of gap
solitons in the presence of a defect mode that is due to the total population inversion
at the centre of the sample. The class of interactions we are concerned with here are
those that produce a light stopping and release operation which can be exploited in order
to achieve a memory cell with high-speed switching at a moderate consumption of
the laser power.

We have previously shown that it is possible to trap a gap soliton on the potential
of Fig. 1 \cite{ref19,ref21}. Figure 2 shows a contour plot in the $(x,t)$-plane, depicting the
local inversion and the propagation and subsequent trapping of a $2\pi$ pulse.
We call the GS that is stopped and oscillating at such defect an information bit. It is important
to note that only such stopped solitons can be used to switch subsequent pulses providing
either true memory readout, higher-density storage, or limiting action so providing
a secure readout. This trapping represents a basic optical memory whereby the GS is
stored in the RPhC. However, in order to demonstrate its usefulness we need to investigate
the read-out or release of information from such a memory device.

In Figs. 3a and 3b we show that a second $2\pi$ gap soliton can be used to release
the stored pulse and effectively read out the optical memory. The soliton readout happens
owing to the collision of the out-of-phase pulse with the trapping area. Our simulations
show that the soliton can be released by the readout pulse as long as the readout GS is
timed to arrive when the trapped soliton is at its maximum deviation towards the front
side of the defect. As the readout soliton arrives it interacts with the trapped pulse
and is dragged into the defect. The trapped and readout solitons cross the center of
the defect together and have sufficient energy to overcome the potential and are released.
Naturally, the released pulse is a higher order breather (see Fig. 3c) and it propagates with
a higher velocity through the resonant photonic crystal.

\begin{figure}
\includegraphics[viewport=100 290 520 650,width=.9\columnwidth,clip]
{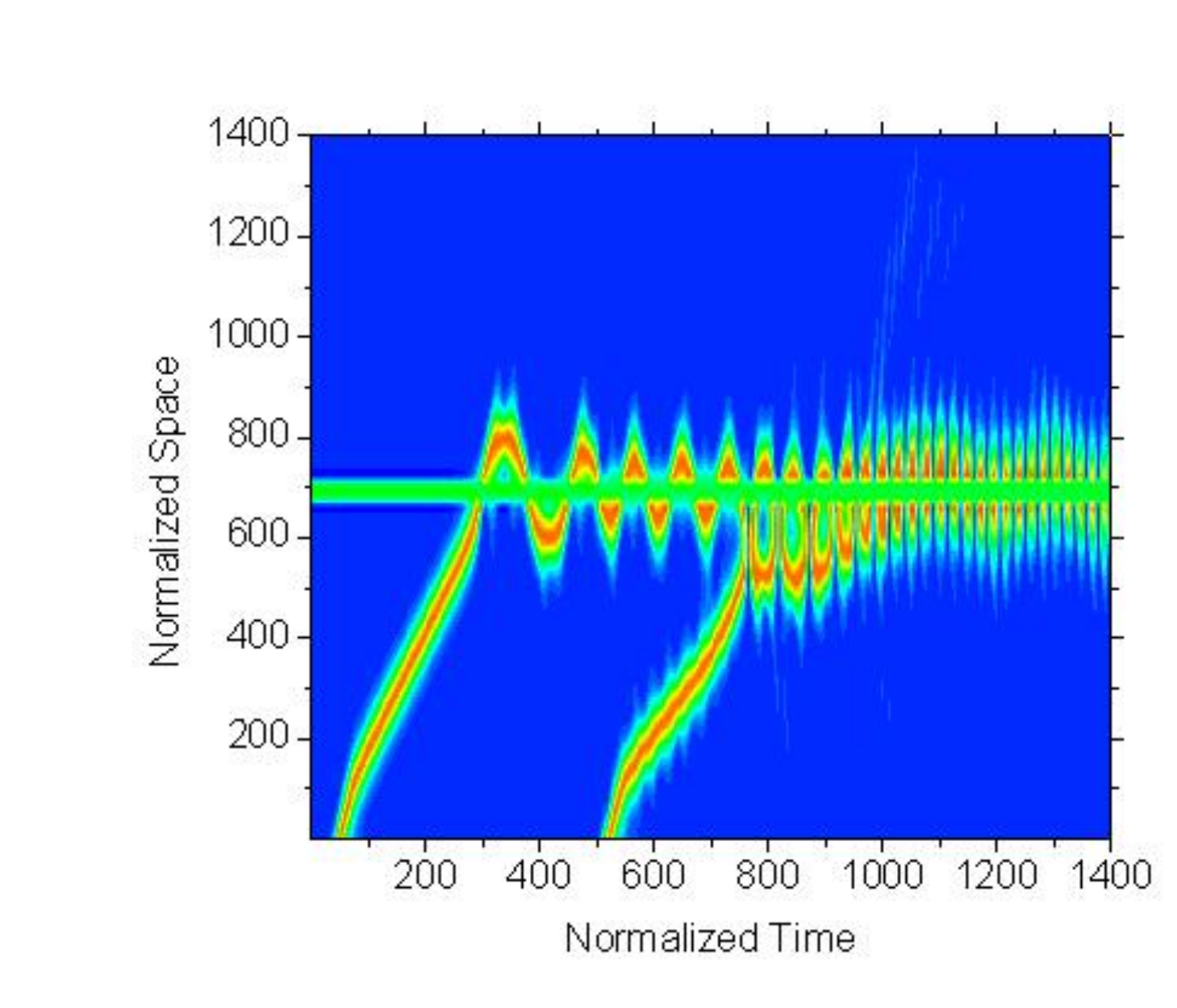}
\caption{ Merger of the readout pulse with the stored one and their
pinning by the defect in the form of quasi-symmetric bound state;
time $t_2 -t_I = 475$. In this case pulses collide on the defect after
5 oscillations of the leading pulse around the defect. }
\label{fig4}
\end{figure}

From this argument it is clear that if the readout GS is not synchronized with the trapped
soliton then they can combine at the defect and can both be trapped. We illustrate this
effect in Fig. 4, where the out-of-phase soliton used to trigger the read-out arrives while
the trapped one is at its turning point on the opposite side of the defect. The resulting
interaction causes the solitons to arrive at the center of the inversion region
from opposite sides.
The result is that the solitons combine to form a trapped mode at the inversion.
This corresponds to a higher order trapped mode as can be seen from the higher frequency
structure on the soliton envelope.
In order to implement a practical optical memory it is essential that the solitons
be timed correctly. This implies that the strength of the inversion in the photonic crystal
should be such that the period of oscillation of the trapped mode is the same as the bit
rate of the data stream. If this condition is not satisfied, the merging breather-like
state turns out to be unstable, dissociates into a soliton trapped to the inversion and
the read-out pulse rejected backward - a read-out failure. This observation
(which is not shown in Figures) is extremely important from the information security
standpoint. That is, since the oscillation period and amplitude determine the open time of
the memory cell a pulse with the wrong phase cannot take away the information bit.

Since the process of the memory read-out is due to the energy imbalance between interacting
solitons and the trapping defect, it is natural to assume that trapping and release of a
gap soliton may also be controlled by means of its interaction with an in-phase, and more
intense pulse. As a relevant example, Fig. 5 shows the results of numerical integration
of the Maxwell-Bloch equations, Eqs.~(\ref{eq4a})--(\ref{eq4c})
for the case of interaction of two solitons with different input amplitudes.
Above a certain threshold, the trailing GS with larger energy permits
the escape of the GS from the trapping area (Fig. 5a). This is due to the repulsive action
of the read-out pulse that is stronger than the trapping action provided by the defect.
What might be interesting from a potential device feasibility stand-point is that the
released soliton emerges the RPhC in the backward direction. Again, this happens only for
a certain bit rate of the data stream. The readout failure is demonstrated in Fig. 5b and 5c.
Here, the speed of the released light is so low that is practically creates another defect
inside the crystal so rendering the original one available for the information storage.

The other important finding that follows from Fig. 5 is the
existence of a stabilization of a higher-order solution to the
perturbed sine-Gordon equation, Eq. (\ref{eq8}), due to the presence
of a local perturbation. This is in sharp contrast to the well-known
fact of the breakup of the $2\pi n$-pulse (with $n$ being an
integer) of the self-induced transparency into a train of isolated
$2\pi$ pulses with different amplitudes, speeds, and durations
\cite{ref24}.

We conclude this Section with indication of an alternative way of stopping light
inside a RPhC. Contrary to the initial conditions of Eqs. (\ref{eq5a})--(\ref{eq5e})
let us now assume that the gain (or total inversion of the atoms) is not squeezed in
the middle of the sample but is distributed from the input facet at certain distance
inside. Using initial conditions pertinent to spontaneous decay \cite{ref25,ref26},
also leads to interesting results. This is summarized in Fig. 6.
The localized field oscillating around the physical boundary between
the pumped and unpumped parts of the RPhC is clearly visible.

\begin{figure}
\includegraphics[viewport=100 320 500 640,width=.9\columnwidth,clip]
{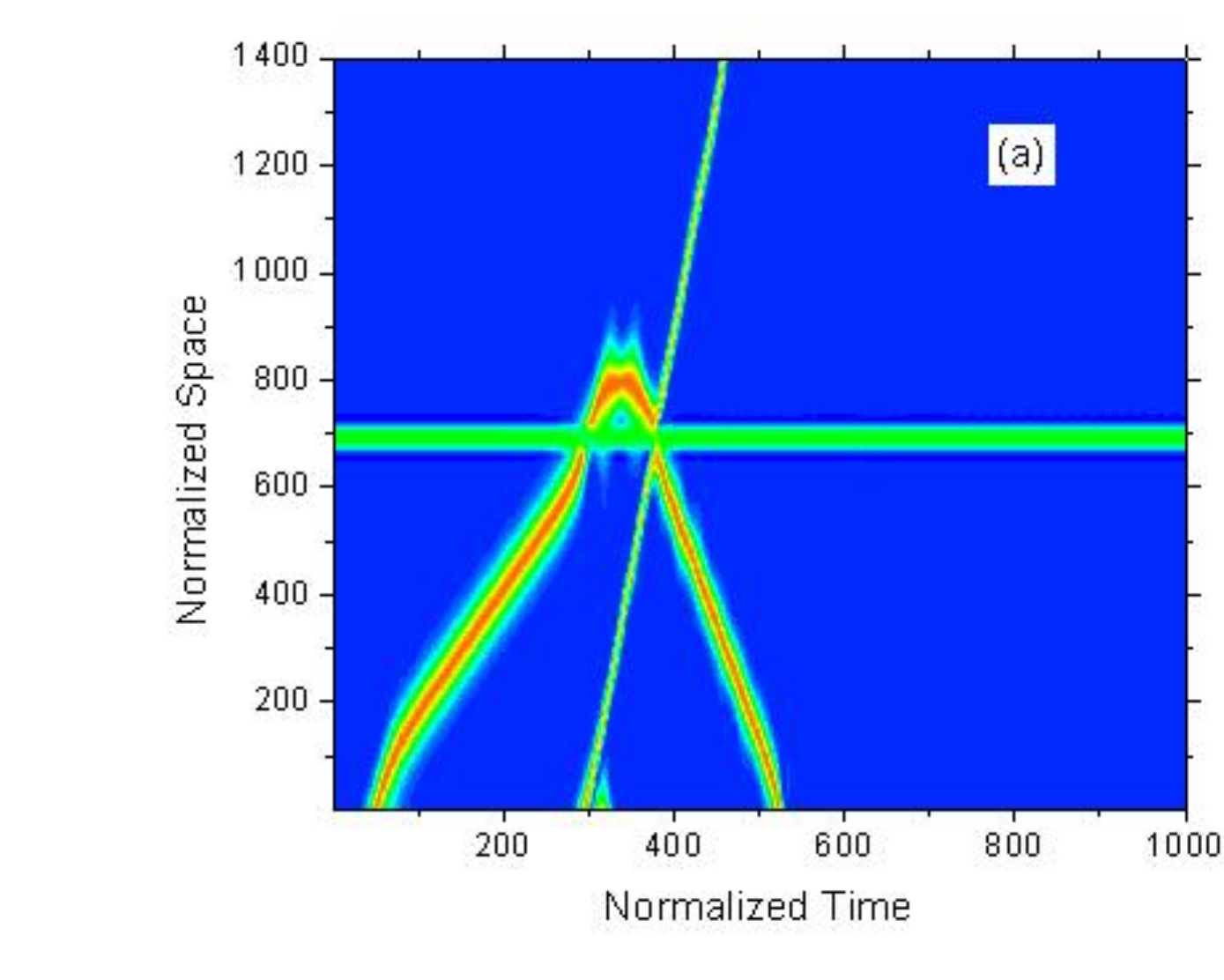}
\includegraphics[viewport=100 320 500 640,width=.9\columnwidth,clip]
{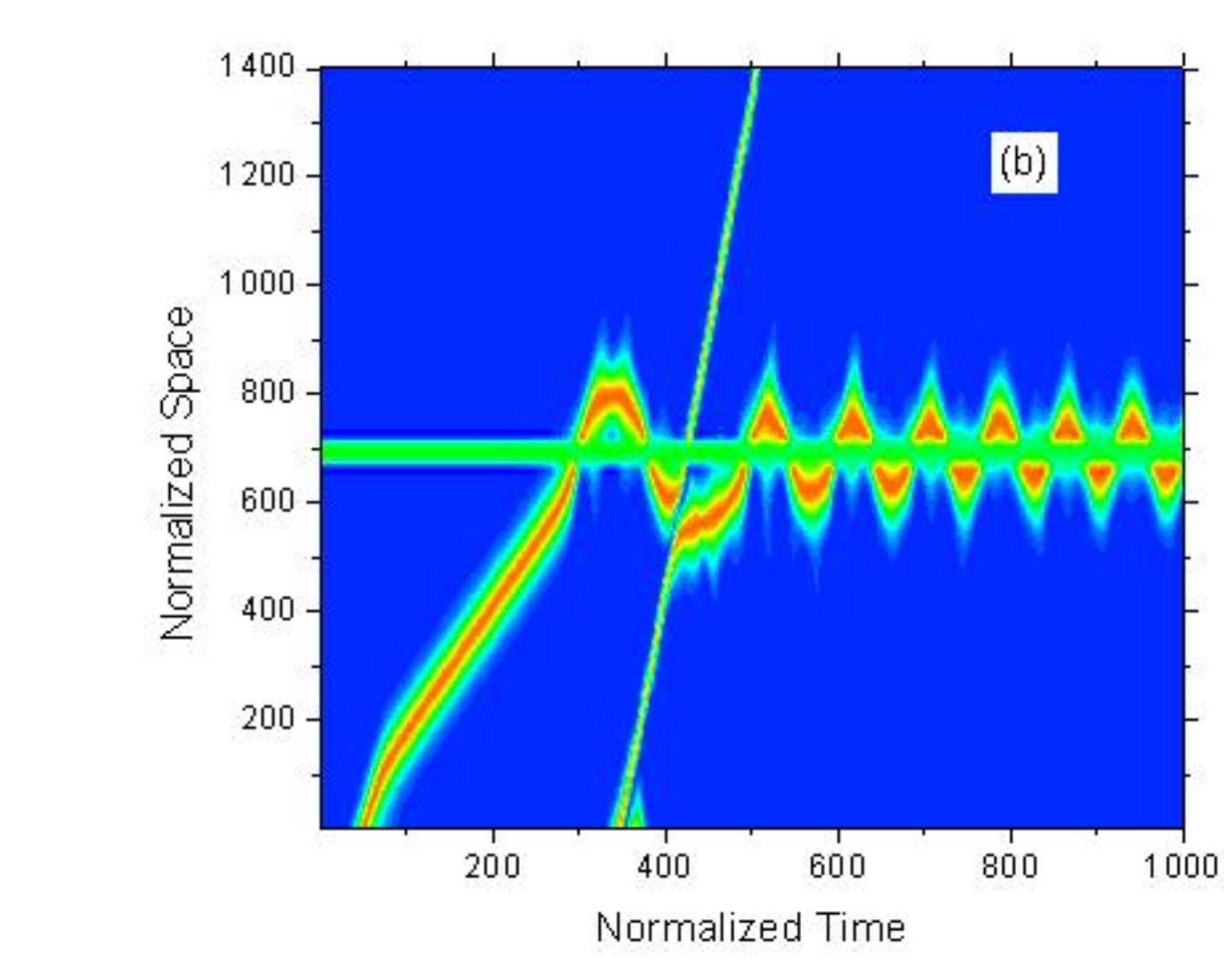}
\caption{ Release of a stored optical pulse by in-phase intense
pulse (a) and readout failure (b); $\Omega_1= 2\Omega_2 /3, t_2 -
t_I = 250$ (a), $t_2 - t_I = 275$ (b). }
\label{fig5}
\end{figure}

\begin{figure}
\includegraphics[viewport=100 320 500 640,width=.9\columnwidth,clip]
{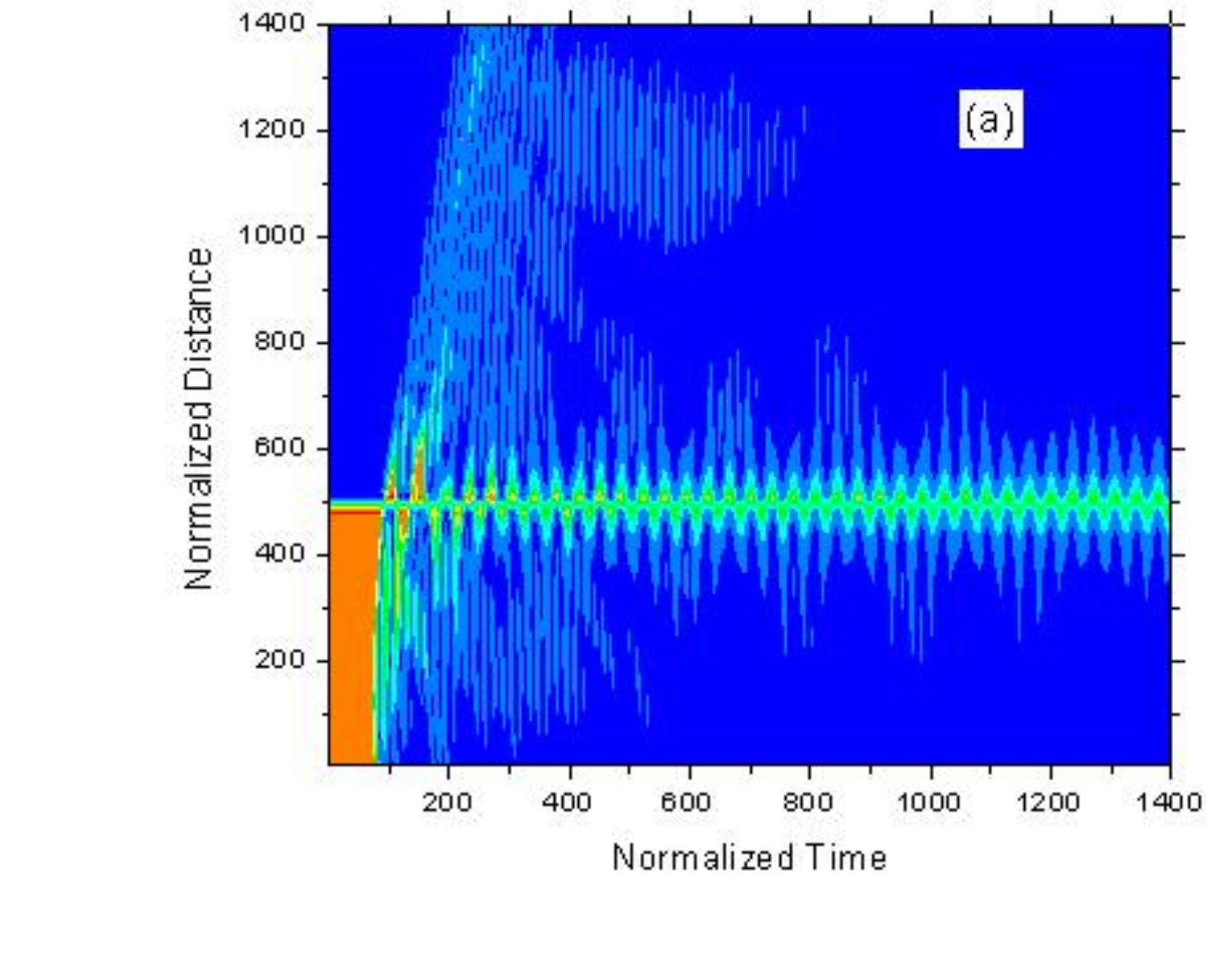}
\includegraphics[viewport=100 320 500 640,width=.9\columnwidth,clip]
{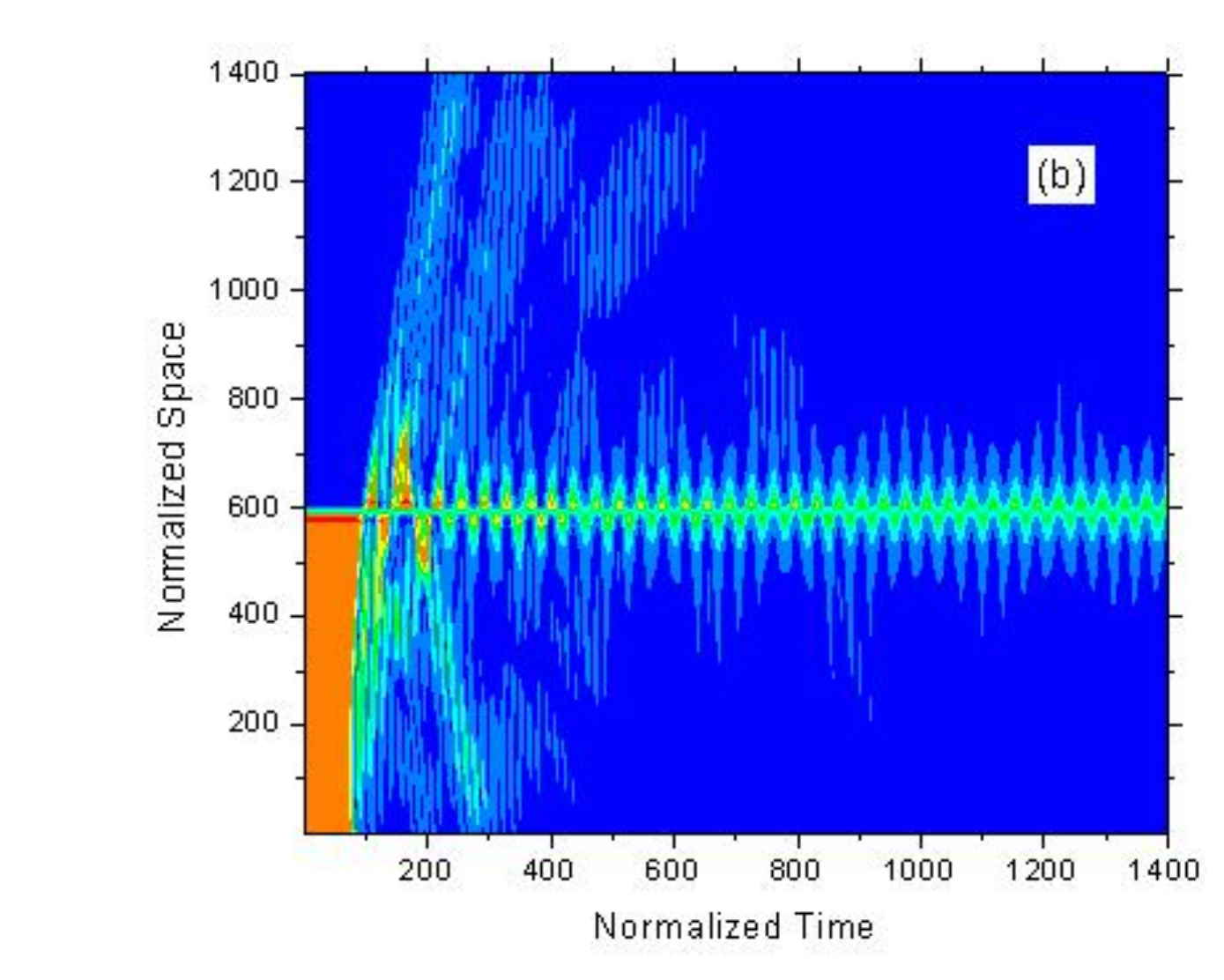}
\includegraphics[viewport=100 320 500 640,width=.9\columnwidth,clip]
{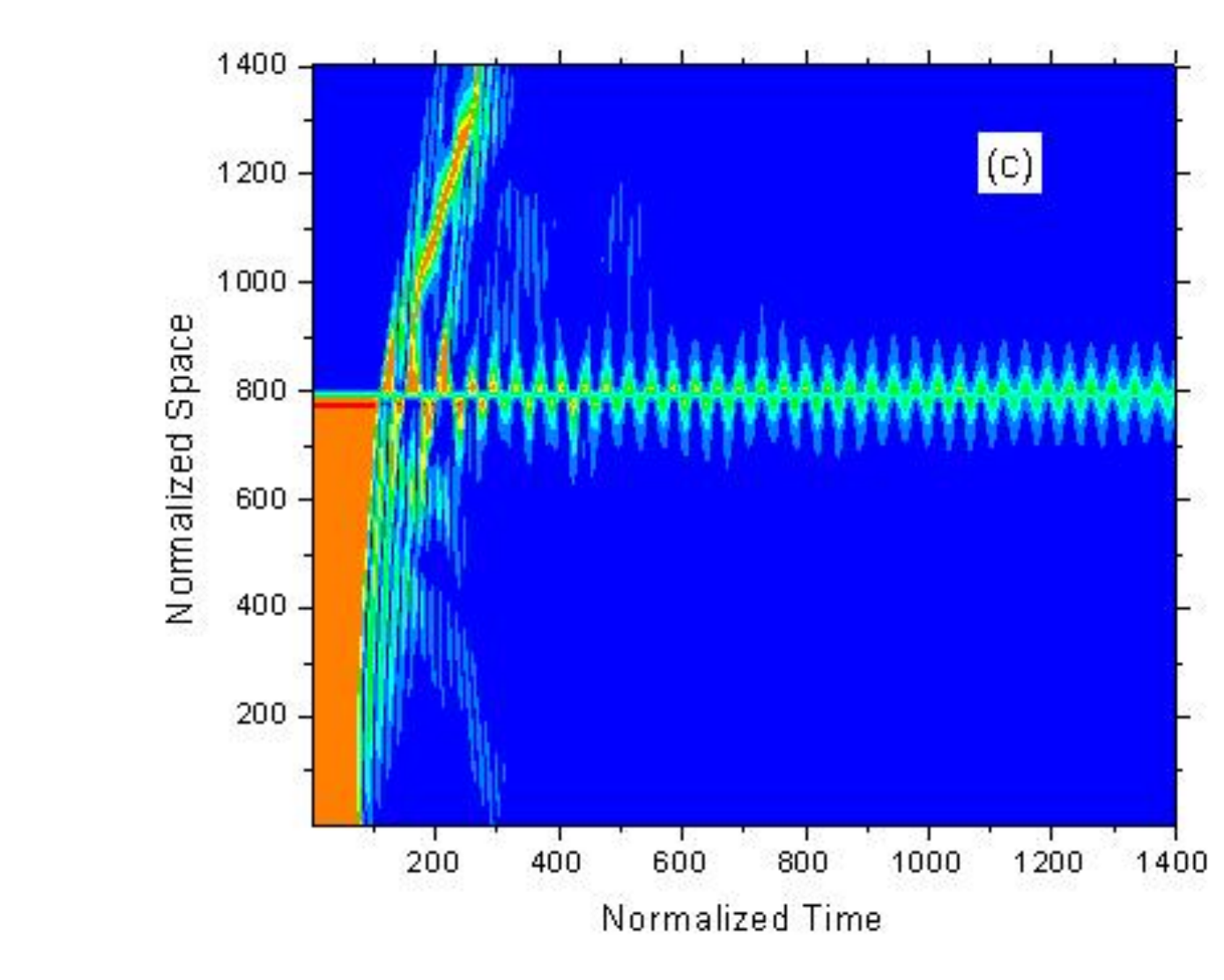}
\caption{
Spontaneous decay and oscillating defect formation at the interface
between pumped and ground-state parts of the RPhC;
the normalized length of the pumped section is 500 (a), 600 (b), and 800 (c).
}
\label{fig6}
\end{figure}

\section{Conclusions}
\label{discuss}

The results presented in this paper are based on studying gap solitons in resonance
photonic crystals.
However, the sine-Gordon equation can also be used to describe a broad and diverse
range of physical systems; for example, the dissociation of the breather solution of
the sine-Gordon equation that is caused by a perturbation of a long Josephson junction
\cite{ref27,ref28,ref29,ref30,ref31,ref32}.
The problem of the stability of the perturbed sine-Gordon equation also results in the
ratchet effect \cite{ref33,ref34}, the high-frequency field interaction with electron
plasmas in a semiconductor superlattice \cite{ref35}, and the spatial localization in
a nonlinear chain \cite{ref36}.
We would therefore expect that similar effects of energy storage and release by
a defect will also be present in these systems.

In conclusion, we have demonstrated that the oscillating immobile
gap solitons created by the presence of a localized region of the
population inversion (or gain) inside a resonance photonic crystal
can be manipulated by a proper choice of bit rate, phase and amplitude
modulation. Developing this idea, we are able to obtain qualitatively
different regimes of the RPhC operation.
A noteworthy observation is that both the delay time and
amplitude difference must exceed a certain level to ensure effective
control over the entire soliton dynamics and its speed, in particular.
A modification of the defect that accomplishes a soliton slowing-down
and trapping can make the dynamics of soliton trains in the resonance
photonic crystal with defects even more interesting.
This comprises a subject of our future work.

\section{Acknowledgements}

We acknowledge helpful discussions with J. Band, S. Flach, W. Hoyer, M. Kira,
G. Kurizki, T. Meier, T. Stroucken, and M. I. Tribelsky. This work was supported in
parts by National Science and Engineering Research Council of Canada,
Photonics Research Ontario, and Max-Planck-Gesellschaft.

\end{document}